\documentclass[12pt]{article}
\usepackage{amssymb,amsmath,epsfig}

\begin{document}

\title{\bf Effects of $f(R)$ Dark Energy on Dissipative Anisotropic Collapsing Fluid}

\author{M. Sharif \thanks{msharif@math.pu.edu.pk} and H. Rizwana
Kausar
\thanks{rizwa\_math@yahoo.com}\\
Department of Mathematics, University of the Punjab,\\
Quaid-e-Azam Campus, Lahore-54590, Pakistan.}

\date{}
\maketitle

\begin{abstract}
The purpose of this paper is to study the effects of dark energy on
dynamics of the collapsing fluid within the framework of metric
$f(R)$ gravity. The fluid distribution is assumed to be locally
anisotropic and undergoing dissipation in the form of heat flow,
null radiations and shear viscosity. For this purpose, we take
general spherical symmetric spacetime. Dynamical equations are
obtained and also some special solutions are found by considering
shearing expansionfree evolution of the fluid. It is found that dark
energy affects the mass of the collapsing matter and rate of
collapse but does not affect the hydrostatic equilibrium.\\
\end{abstract}

\section{Introduction}

Dark energy (DE) and gravitational collapse are the two noteworthy
issues of cosmology and gravitational physics. Recent
observational data \cite{super1}-\cite{sneIa} indicate that our
universe is expanding. This acceleration is explained in terms of
DE, which may be explained in modified gravity models. On the
other hand, gravitational collapse is the basic process driving
evolution within galaxies, assembling giant molecular clouds and
producing stars.

When the Einstein-Hilbert (EH) gravitational action in General
Relativity (GR),
\begin{equation}\label{1}
S_{EH}=\frac{1}{2 \kappa}\int d^{4}x\sqrt{-g}R,
\end{equation}
is re-written in the modified form as follows
\begin{equation}\label{2}
S_{modif}=\frac{1}{2\kappa}\int d^{4}x\sqrt{-g}f(R),
\end{equation}
the presence of $f(R)$ function may be understood as the
introduction of an effective fluid which is not restricted to hold
the usual energy conditions. An important feature of this theory is
that the modified field equations can be written in the form of
Einstein tensor which makes it easy to compare with GR. This is done
by taking all the higher order corrections to the curvature on the
right hand side of the field equations and defining it as a "Dark
source" term or "curvature fluid". In the following, the field
equations in the metric approach are obtained as
\begin{equation}\label{3}
F(R)R_{\alpha\beta}-\frac{1}{2}f(R)g_{\alpha\beta}-\nabla_{\alpha}
\nabla_{\beta}F(R)+ g_{\alpha\beta} \Box F(R)=\kappa
T_{\alpha\beta},\quad(\alpha,\beta=0,1,2,3),
\end{equation}
where $F(R)\equiv df(R)/dR$. When we re-write this equation in the
above mentioned form, it follows that
\begin{equation}\label{4}
G_{\alpha\beta}=\frac{\kappa}{F}(T_{\alpha\beta}^{m}+T_{\alpha\beta}^{D}),
\end{equation}
where
\begin{equation}\label{5}
T_{\alpha\beta}^{D}=\frac{1}{\kappa}\left[\frac{f(R)+R
F(R)}{2}g_{\alpha\beta}+\nabla_{\alpha} \nabla_{\beta}F(R)
-g_{\alpha\beta} \Box F(R)\right].
\end{equation}
In this way, DE can be thought of as having the geometrical origin
rather than some additional scalar fields which are added by hand to
the matter part. Therefore, $f(R)$ theory of gravity may be used to
explain the present accelerating expansion of the universe.

This theory has many applications in cosmology and gravity such as
inflation, local gravity constraints, cosmological perturbations
and spherically symmetric solutions in weak and strong
gravitational backgrounds. In the last few years, a considerable
amount of analysis and theoretical observations have been made in
order to compare the preliminary successes of $f(R)$ models with
the great achievements of GR. For example, Capozziello \emph{et
al.} \cite{capo} analyzed the relation between spherical symmetry
and the weak field limit of $f(R)$ theory and compared the results
with GR. In strong gravitational background such as neutron star
and white dwarfs, one needs to take into account the backcreation
of gravitational potentials to the field equations. The structure
of the relativistic stars in $f(R)$ theory has been discussed by
many authors \cite{95}-\cite{99}.

Cai \emph{et al.} \cite{Cai} derived the generalized Misner-Sharp
energy in $f(R)$ gravity for spherically symmetric spacetime. They
found that unlike GR, the existence of the generalized
Misner-Sharp energy depends on a constraint condition. Erickcek
\textit{et al.} \cite{eric} found unique exterior solution for a
stellar object by matching it with interior solution in the
presence of matter source. Kainulainen \textit{et al.} \cite{kain}
studied the interior spacetime of stars in Palatini $f(R)$
gravity. de la Cruz-Dombriz \emph{et al.} \cite{cruz} discussed
the problem of finding static spherically symmetric black hole
solutions in $f(R)$ theory. They explored several aspects of
constant curvature solutions and thermodynamical properties. In a
recent paper \cite{we}, we have investigated perfect fluid
gravitational collapse in this theory and found that constant
scalar curvature term $f(R_0)$ acts as a source of repulsive force
and thus slows down the collapse of matter.

In GR, Oppenheimer and Snyder \cite{OS} innovated the first
mathematical model for the description of gravitational collapse of
stars. After that many approaches are adopted for the physical
description of the fluid in order to form self-gravitating objects.
During fluid evolution, self-gravitating objects may pass through
phases of intense dynamical activities for which quasi-static
approximation is not reliable. For instance, the collapse of very
massive stars \cite{ben}, the quick collapse phase yielding neutron
star formation \cite{myra} and the peculiar stars. The peculiar
stars are very dense, strongly magnetic and are created when massive
stars die by collapse. Many of the cooler chemically peculiar stars
are the result of the mixing of nuclear fusion products from the
interior of the star to its surface.

Misner and Sharp \cite{MS} discussed the gravitational collapse by
taking spherically symmetric ideal fluid. They provided a full
account of the dynamical equations governing the adiabatic
relativistic collapse. Dissipative process plays dominant role in
the formation and evolution of stars. Vaidya \cite{V} introduced
the idea of outgoing radiations in collapse giving non-vacuum
exterior outside the stars. It was physically a quite reasonable
assumption as radiation is a confirmation that dissipative
processes are occurring, causing loss of thermal energy of the
system which is an effective way of decreasing internal pressure.
Cai and Wang \cite{CW} studied the formation of black holes in the
background of DE. Herrera \emph{et al.} \cite{H1} investigated the
dynamics of gravitational collapse which undergoes dissipation in
the form of heat flow and radiation. The same authors \cite{H2}
also provided detailed discussion on the physical meaning of
expansionfree fluid evolution. Di Prisco \emph{et al.} \cite{P}
explored gravitational collapse by adding charge and dissipation
in the form of shear viscosity. The dynamical and transport
equations were coupled to observe the effects of dissipation over
collapsing process. Recently, Sharif \emph{et al.}
\cite{szak}-\cite{sa} explored different aspects of gravitational
collapse by using all the three types of symmetry.

In this paper, we discuss how DE generated by curvature fluid
affects the dynamics of dissipative gravitational collapse. The
format of the paper is as follows. In next section \textbf{2}, we
present spacetimes and energy-momentum tensor for dissipative fluid.
Section \textbf{3} is devoted to formulate the modified field
equations and dynamical equations in $f(R)$ gravity. In section
\textbf{4}, special solutions are discussed. The last section
\textbf{5} concludes the main results of the paper.

\section{Spacetimes and Collapsing Matter}

We take spherical symmetry about an origin $O$ which is divided
into two regions, interior and exterior by $3D$ hypersurface
$\Sigma$ centered at $O$. The interior spacetime to $\Sigma$ can
be represented by the line element
\begin{equation}\setcounter{equation}{1}\label{6}
ds^2=A^2(t,r)dt^{2}-B^2(t,r)dr^{2}-C^2(t,r)(d\theta^{2}+\sin^2\theta
d\phi^{2}).
\end{equation}
For the exterior spacetime to $\Sigma$, we take the Vaidya
spacetime given by the line element
\begin{equation}\label{7}
ds^2=[1-\frac{2m(\nu)}{r}]d\nu^2+2drd\nu-r^2(d\theta^2+\sin^2\theta
d\phi^2),
\end{equation}
where $m(\nu)$ represents the total mass and $\nu$ is the retarded
time. In the interior region, we assume a distribution of
anisotropic collapsing fluid which undergoes a dissipation in the
form of heat flow, null radiations and shearing viscosity. The
energy-momentum tensor with such properties is given by
\begin{equation}\label{8}
T_{\alpha\beta}=(\rho+p_{\perp})u_{\alpha}u_{\beta}-p_{\perp}g_{\alpha\beta}+
(p_r-p_{\perp})\chi_{\alpha}
\chi_{\beta}+u_{\alpha}q_{\beta}+q_{\alpha}u_{\beta}+\epsilon
l_\alpha l_\beta-2\eta\sigma_{\alpha\beta}.
\end{equation}
Here, we have $\rho$ as the energy density, $p_{\perp}$ the
tangential pressure, $p_r$ the radial pressure, $q_{\alpha}$ the
heat flux, $\eta$ the coefficient of shear viscosity, $u_{\alpha}$
the four-velocity of the fluid, $\chi_{\alpha}$ the unit
four-vector along the radial direction, $l_\alpha$ a radial null
four-vector and $\epsilon$ the energy density of the null fluid
describing dissipation in the free streaming approximation. These
quantities satisfy the relations
\begin{eqnarray}\nonumber
&&u^{\alpha}u_{\alpha}=1,\quad\chi^{\alpha}\chi_{\alpha}=-1,\quad
l^{\alpha}u_{\alpha}=1,\\\label{9} &&u^{\alpha}q_{\alpha}=0,\quad
\chi^{\alpha}u_{\alpha}=0,\quad l^{\alpha}l_{\alpha}=0
\end{eqnarray}
which are obtained from the following definitions in co-moving
coordinates
\begin{eqnarray}\label{10}
u^{\alpha}=A^{-1}\delta^{\alpha}_{0},\quad
\chi^{\alpha}=B^{-1}\delta^{\alpha}_{1},\quad
q^{a}=qB^{-1}\delta^{\alpha}_{1},\quad
l^{a}=A^{-1}\delta^{\alpha}_{0}+B^{-1}\delta^{\alpha}_{1}.
\end{eqnarray}
Here $q$ is a function of $t$ and $r$. The shear tensor
$\sigma_{ab}$ is defined by
\begin{equation}\label{11}
\sigma_{\alpha\beta}=u_{(\alpha;\beta)}-a_{(\alpha}u_{\beta)}-\frac{1}{3}\Theta
(g_{\alpha\beta}-u_{\alpha}u_{\beta}),
\end{equation}
where the acceleration $a_{a}$ and the expansion $\Theta$ are given
by
\begin{equation}\label{12}
a_{\alpha}=u_{\alpha;\beta}u^{\beta},\quad
\Theta=u^{\alpha}_{;\alpha}.
\end{equation}
The bulk viscosity does not appear explicitly as it has been
absorbed in the form of radial and tangential pressures of the
collapsing fluid. From Eqs.(\ref{10}) and (\ref{11}), the non-zero
components of the shear tensor are
\begin{equation}\label{13}
\sigma_{11}=-\frac{2}{3}{B^{2}}\sigma,\quad
\sigma_{22}=\frac{1}{3}{C^{2}}\sigma,\quad
\sigma_{33}=\sigma_{22}\sin^2\theta.
\end{equation}
The shear scalar $\sigma$ is given by
\begin{equation}\label{14}
\sigma=\frac{1}{A}\left(\frac{\dot{B}}{B}-\frac{\dot{C}}{C}\right).
\end{equation}
Using Eqs.(\ref{10}) and (\ref{12}), it follows that
\begin{equation}\label{15}
a_{1}=-\frac{A'}{A},\quad a^2=a^\alpha
a_\alpha=\left(\frac{A'}{AB}\right)^2,\quad
\Theta=\frac{1}{A}\left(\frac{\dot{B}}{B}+2\frac{\dot{C}}{C
}\right),
\end{equation}
where dot and prime represent derivative with respect to $t$ and $r$
respectively.

\section{The Field Equations and the Dynamical Equations in f(R) Gravity}

The field equations (\ref{3}) for the interior metric take the
following form
\begin{eqnarray}\setcounter{equation}{1}\nonumber
\frac{AA''}{B^2}-\frac{\ddot{B}}{B}+\frac{\dot{A}\dot{B}}{AB}-\frac{AA'B'}{B^3}
-\frac{2\ddot{C}}{C}
+\frac{2\dot{A}\dot{C}}{AC}+\frac{2AA'C'}{B^2C}-\frac{A^2}{2}\frac{f(R)}{F}&&\\\label{16}
-\frac{A^2{F''}}{B^2F}-\frac{\dot{F}}{F}(\frac{-\dot{B}}{B}+\frac{2\dot{C}}{C})
-\frac{F'A^2}{F B^2}(\frac{-B'}{B}+\frac{2C'}{C})
=8{\pi}({\rho}+\epsilon)A^{2},&&\\\label{17}
-2(\frac{\dot{C'}}{C}-\frac{\dot{C}A'}{CA}-\frac{\dot{B}C'}{BC})-\frac{\dot{F'}}{
F}+\frac{A'\dot{F}}{AF}+\frac{\dot{B}F'}{BF}=-8{\pi}(q+\epsilon)AB,&&\\\nonumber
\frac{-A''}{A}+\frac{B\ddot{B}}{A^2}-\frac{\dot{A}\dot{B}B}{A^3}+\frac{A'B'}{AB}-
\frac{2C''}{C}
+\frac{2\dot{C}\dot{B}B}{A^2C}+\frac{2B'C'}{BC}&&\\\nonumber
+\frac{B^2}{2}\frac{f(R)}{F}
-\frac{B^2\ddot{F}}{A^2F}+\frac{\dot{F}B^2}{FA^2}(\frac{\dot{A}}{A}+\frac{2\dot{C}}{C})
+\frac{F'}{F}(\frac{A'}{A}+\frac{2C'}{C})&&\\\label{18}
=8{\pi}(p_r+\epsilon+\frac{4}{3}{\eta}{\sigma})B^{2},&&\\\nonumber
\frac{\ddot{C}}{CA^2}-\frac{\dot{A}\dot{C}}{CA^3}-\frac{A'C'}{B^2AC}-\frac{C''}{B^2C}
+\frac{\dot{C}\dot{B}}{A^2BC}
+\frac{B'C'}{B^3C}+\frac{1}{C^2}&&\\\nonumber
+\frac{\dot{C}^2}{A^2C^2}-\frac{C'^2}{B^2C^2}
+\frac{C^2}{2}\frac{f(R)}{F}-\frac{\ddot{F}}{A^2}+\frac{F''}{B^2}
+\frac{\dot{F}}{A^2}(\frac{\dot{A}}{A}-\frac{\dot{B}}{B}+\frac{\dot{C}}{C})
\\\label{19}+\frac{F'}{B^2}(\frac{A'}{A}-\frac{B'}{B}+\frac{C'}{C})
=8{\pi}(p_{\perp}-\frac{2}{3}{\eta}{\sigma}).
\end{eqnarray}
In the most general case, we have nine variables with four
equations. One cannot solve this system of equations unless some
assumptions are imposed. An expansionfree motion of the system,
i.e., $\Theta=0$, will be used in section 4 which may lead to some
interesting results.

In order to develop dynamical equations that help to study the
properties of collapsing process, we shall use Misner and Sharp
\cite{MS} formalism. The mass function is defined by
\begin{equation}\label{20}
M=\frac{C}{2}(1+g^{\mu\nu}C_{,\mu}C_{,\nu})=\frac{C}{2}\left(1+\frac{\dot{C}^2}{A^2}
-\frac{C'^2}{B^2}\right).
\end{equation}
The proper time and radial derivatives are given by
\begin{equation}\label{45}
D_{T}=\frac{1}{A}\frac{\partial}{\partial t},\quad
D_{C}=\frac{1}{C'}\frac{\partial}{\partial r},
\end{equation}
where $C$ is the areal radius of a spherical surface inside the
boundary. The velocity of the collapsing fluid is defined by the
proper time derivative of $C$, i.e.,
\begin{equation}\label{22}
U=D_{T}C=\frac{\dot{C}}{A},
\end{equation}
which is always negative. Using this expression, Eq.(\ref{20})
implies that
\begin{equation}\label{23}
E\equiv\frac{C'}{B}=[1+U^{2}+\frac{2M}{C}]^{1/2}.
\end{equation}
When we make use of Eqs.(\ref{14}),(\ref{15}) and (\ref{45}) in
Eq.(\ref{17}), we obtain
\begin{eqnarray}\nonumber
&&E[\frac{1}{3}D_{C}(\Theta-\sigma)-\frac{\sigma}{C}]\\\label{24}
&&=\frac{4\pi}{F}\left[(q+\epsilon)\frac{C'}{E} +D_T
F'-\frac{UC'}{\dot{C}}D_C AD_T F-\frac{EUF'}{\dot{C}}D_C B\right].
\end{eqnarray}

The rate of change of mass in Eq.(\ref{20}) with respect to proper
time, with the use of Eqs.(\ref{16})-(\ref{19}), is given by
\begin{eqnarray}\nonumber
D_{T}M&=&\frac{C^2}{F}\left[-4\pi\{(\rho+2\epsilon+p_r-2p_{\perp}
+\frac{8}{3}{\eta}{\sigma})U+E(q+\epsilon)\}\right.\\\nonumber
&-&U\left\{\frac{F}{2C^2}-\frac{FE^2}{C^2}+\frac{FD_TU}{C}+\frac{E^2}{C'^2}
\{\frac{\dot{F}C'}{C}+\frac{3F''}{2}E\dot{F}D_CB\}\right.\\\nonumber&+&\left.
\frac{ED_AD_TF}{2\dot{C}}+\frac{f(R)}{2}\right\}
-U^3\left\{\frac{F}{C^2}+\frac{\ddot{F}}{\dot{C}^2}+\frac{\dot{F}}{\dot{C}}
(\frac{D_TA}{2\dot{C}}-\frac{1}{C})\right\}\\\nonumber&+&U^2\left\{(\frac{F}{C}
-\frac{F'}{2C'})D_CA
+\frac{3\dot{F}D_TB}{2EC'}\right\}\frac{E^2}{\dot{C}}\\\label{25}
&-&\left.\frac{E^3}{2C'}[D_TF'+ED_CFD_TB]\right].
\end{eqnarray}
This represents variation of total energy inside a collapsing
surface of radius $C$. The first two terms inside the square
brackets have negative signs which show that the total energy is
being dissipated during collapse. However, in the case of collapse
$U<0$, the first term $(\rho+2\epsilon+p_r-2p_{\perp}
+\frac{8}{3}{\eta}{\sigma})$ increases the energy density through
the rate of work being done by the effective anisotropic pressure
and radiation density of the null fluid. Here we may use equation
of state to change energy density into pressure. The second term
$E(q+\epsilon)$ has negative sign which describes that energy is
leaving the system due to heat flux and radiations. All other
terms show the contribution of the DE in the form of function
$f(R)$ and its derivatives. We know that DE exerts a repulsive
force on its surrounding thus we may conclude from the above
expression that DE reduces the mass of the collapsing matter due
to its negative pressure. It is mentioned here that in GR, only
the first two terms excluding energy density $\rho$ and tangential
pressure appear.

Similarly, we can calculate
\begin{eqnarray}\nonumber
D_{C}M&=&\frac{C^2}{2F}\left[8\pi\{\rho+2\epsilon+p_r+\frac{4}{3}{\eta}{\sigma}
+\frac{U}{E}(q+\epsilon)\}\right.\\\nonumber
&+&U^2\left(\frac{F}{C}-D_TFD_CA\right)
-U\left\{ED_CF\left(\frac{D_TB}{C'}
-\frac{ED_CA}{\dot{C}}\right)\right.\\\nonumber&+&
\left.\frac{EFD_CA}{CC'}\right\}+D_{TT}F
-\frac{F''E^2}{C'^2}-\frac{ED_TF'}{C'}
-\frac{E}{C'}(D_TFD_TB\\\label{26} &+&
E^2D_CFD_CB)+\left.\frac{FE^2}{C^2}+\frac{F}{C^2}+\frac{2FD_TU}{C}\right].
\end{eqnarray}
This equation describes how different quantities influence the
mass between neighboring surfaces of radius $C$ in the fluid
distribution. The first two terms and their description in the
above expression are almost the same as in GR \cite{H2} except for
the factor $(p_r+\frac{4}{3}{\eta})$. The appearance of this
factor is due to the complicated field equations. The remaining
terms represent contribution of DE due to curvature fluid. Taking
integral of Eq.(\ref{26}) over $C$, we have
\begin{eqnarray}\nonumber
M&=&\frac{1}{2}\int^{C}_{0}\frac{C^2}{2F}\left[8\pi\{\rho+2\epsilon+p_r+
\frac{4}{3}{\eta}{\sigma}
+\frac{U}{E}(q+\epsilon)\}\right.\\\nonumber
&+&U^2\left(\frac{F}{C}-D_TFD_CA\right)
-U\left\{ED_CF\left(\frac{D_TB}{C'}
-\frac{ED_CA}{\dot{C}}\right)\right.\\\nonumber&+&
\left.\frac{EFD_CA}{CC'}\right\}+D_{TT}F
-\frac{F''E^2}{C'^2}-\frac{ED_TF'}{C'}
-\frac{E}{C'}(D_TFD_TB\\\label{27} &+&
E^2D_CFD_CB)+\left.\frac{FE^2}{C^2}+\frac{F}{C^2}+\frac{2FD_TU}{C}\right]dC.
\end{eqnarray}

The dynamical equations can be obtained from the contracted Bianchi
identities. Consider the following two equations
\begin{eqnarray}\label{52}
T^{\alpha\beta}_{;\beta}u_{\alpha}=0,\quad T^{\alpha\beta}_{;\beta}
\chi_{\alpha}=0
\end{eqnarray}
which yield
\begin{eqnarray}\nonumber
&&\frac{1}{A}\left[(\rho+\epsilon)^{\cdot}+(\rho+2\epsilon+p_r+\frac{4}{3}\eta
F)\frac{\dot{B}}{B}+2(\rho+\epsilon+p_{\perp}-\frac{2}{3}\eta
\sigma)\frac{\dot{C}}{C}\right]\\\label{28}
&&+\frac{1}{B}\left[(q+\epsilon)'+2(q+\epsilon)\frac{(AC)'}{AC}\right]=0,\\\nonumber
&&\frac{-1}{A}\left[(q+\epsilon)^{\cdot}+2(q+\epsilon)(\frac{\dot{B}}{B}
+\frac{\dot{C}}{C})\right]
-\frac{1}{B}\left[(p_r+\epsilon+\frac{4}{3}\eta\sigma)'\right.\\\label{29}
&&\left.+(\rho+p_r+2\epsilon+\frac{4}{3}\eta
\sigma)\frac{A'}{A}+2(p_r-p_{\perp}+\epsilon+2\eta
\sigma)\frac{C'}{C}\right]=0.
\end{eqnarray}
Using Eqs.(\ref{14}), (\ref{15}), (\ref{45}) and (\ref{23}), it
follows that
\begin{eqnarray}\nonumber
&&D_T(\rho+\epsilon)+\frac{1}{3}(3\rho+4\epsilon+p_r+2p_{\perp})\Theta+\frac{2}{3}
(\epsilon+p_r-p_{\perp}-2\eta \sigma)\sigma\\\label{30}
&&+ED_C(q+\epsilon)+2(q+\epsilon)(a+\frac{E}{C})=0,\\\nonumber
&&D_T(q+\epsilon)+\frac{2}{3}(q+\epsilon)(2\Theta+\sigma)+ED_C(p_r+\epsilon
+\frac{4}{3}\eta\sigma)+(\rho+p_r\\\label{31}
&&+2\epsilon+\frac{4}{3}\eta
\sigma)a+2(p_r-p_{\perp}+\epsilon+2\eta \sigma)\frac{E}{C}=0.
\end{eqnarray}
The acceleration $D_{T}U$ of the collapsing matter inside the
hypersurface is obtained by using Eqs.(\ref{19})-(\ref{45}) and
(\ref{23})
\begin{eqnarray}\nonumber
D_{T}U&=&\frac{2M}{C^2}+\frac{8\pi
C}{F}(p_{\perp}-\frac{2}{3}\eta\sigma)+Ea+ED_CE-\frac{UED_TB}{C'}\\\nonumber
&-&\frac{C f(R)}{2F}
-\frac{C}{F}\{-D_{TT}F+ED_C(\frac{F'}{B})-\sigma D_TF\\\label{32}
&+&E^2(\frac{UD_CA}{\dot{C}}+\frac{1}{C})D_CF \}.
\end{eqnarray}
Substituting $a$ from Eq.(\ref{32}) into (\ref{31}), it follows
that
\begin{eqnarray}\nonumber
&&(\rho+p_r+2\epsilon+\frac{4}{3}\eta
\sigma)D_TU=(\rho+p_r+2\epsilon+\frac{4}{3}\eta
\sigma)\left[\frac{8\pi
C}{F}(p_{\perp}-\frac{2}{3}\eta\sigma)\right.\\\nonumber
&&\left.+\frac{2M}{C^2}- \frac{C
f(R)}{2F}+\frac{C}{F}D_{TT}F-\sigma D_TF\right]-
E^{2}\left[D_C(p_r+\epsilon+\frac{4}{3}\eta
\sigma)+\frac{2}{C}(p_r\right.\\\nonumber
&&\left.-p_{\perp}+\epsilon+2\eta
\sigma)+(\rho+p_r+2\epsilon+\frac{4}{3}\eta\sigma)
\left(\frac{UD_CA}{\dot{C}}+\frac{1}{C}\right)\frac{CD_CF}{F}\right]
\\\nonumber&&-E\left[D_T(q+\epsilon)+2(q+\epsilon)(\frac{2U}{C}+\sigma)
+(\rho+p_r+2\epsilon+\frac{4}{3}\eta
\sigma)\right.\\\label{33}&&\left.\times
\left(-D_CE+\frac{UD_TB}{C'}+\frac{C}{F}D_C(\frac{F'}{B})\right)\right].
\end{eqnarray}
This shows the role of different forces on the collapsing process.
The term within the brackets on the left hand side stands for
"effective" inertial mass and the remaining term is acceleration.
The first term on the right hand side represents gravitational
force (the passive gravitational mass by equivalence mass). The
term within the first square brackets shows how dissipative terms
and DE affect the passive gravitational mass. The first two terms
in the second square brackets are gradient of the effective
pressure and effect of local anisotropy of pressure with negative
sign which increases the rate of collapse. While the other term
shows the contribution of DE collective with effective pressure.
Here the presence of $U$ (with negative sign) indicates that this
term slows down the rate of collapse. The last square brackets
depicts the combine role of each type of matter component.

\section{Some Special Solutions}

\subsection{Shearing Expansionfree Dissipative Fluid}

Now we use the condition of expansionfree motion, i.e., $\Theta=0$,
to find some solutions. Using this condition, Eq.(\ref{15}) yields
\begin{equation}\setcounter{equation}{1}\label{34}
\frac{\dot{B}}{B}=-2\frac{\dot{C}}{C}.
\end{equation}
The physical meaning of this condition is discussed with detail in
\cite{H2}. On integration, we get
\begin{equation}\label{35}
B=\frac{g_1(r)}{C^2}.
\end{equation}
where $g_1(r)$ is an arbitrary function. Substituting Eq.(4.1) in
(\ref{17}), we get
\begin{equation}\label{36}
2(\frac{\dot{C'}}{\dot{C}}-\frac{A'}{A}-2\frac{C'}{C})+(\frac{\dot{F'}}{
F}-\frac{A'\dot{F}}{AF})\frac{C}{\dot{C}}+2\frac{F'}{F}=8{\pi}(q+\epsilon)
\frac{AB}{F}\frac{C}{\dot{C}}
\end{equation}
which, by integrating, gives
\begin{equation}\label{37}
A=\frac{F^2\dot{C}C^2}{\tau(t)}e^{-\int{(8{\pi}(q+\epsilon)\frac{Ag_1}{FC^2}
+\frac{\dot{F'}}{ F}-\frac{A'\dot{F}}{AF})\frac{C}{\dot{C}}dr}}.
\end{equation}
Thus the interior metric becomes
\begin{eqnarray}\nonumber
ds^2&=&\left(\frac{F^2\dot{C}C^2}{\tau(t)}\exp[{-\int{(8{\pi}(q+\epsilon)
\frac{Ag_1}{FC^2}+\frac{\dot{F'}}{
F}-\frac{A'\dot{F}}{AF})\frac{C}{\dot{C}}dr}}]\right)^2dt^2
\\\label{38}&&-(\frac{g_1}{C^2})^2dr^2-C^2(t,r)(d\theta^{2}+\sin^2\theta
d\phi^{2}).
\end{eqnarray}
The above metric represents spherically symmetric anisotropic fluid
which is going shearing expansionfree evolution. Assuming the
condition of constant scalar curvature ($R=R_{c}$), according to
which $F(R_c)=constant$, the metric for nondissipative
($q=\epsilon=0$) case is reduced to
\begin{equation}\label{44}
ds^2=\left(\frac{F_c^2\dot{C}C^2}{\tau(t)}\right)^2dt^2-\frac{1}{C^4}dr^2
-C^2(t,r)(d\theta^{2}+\sin^2\theta d\phi^{2}).
\end{equation}
Here we take $g_1(r)=1$. It is mentioned here that $f(R)$ theory, in
the case of constant scalar curvature, exhibits behavior just like
solutions with cosmological constant in GR. This is one of the
reason why the DE issue can be addressed using this theory.
Moreover, in this case, Birkhoff theorem holds, i.e., stationary
solutions are also static \cite{capo} which does not hold for every
$f(R)$.

\subsection{Shearing Expansionfree Nondissipative Perfect Fluid}

Here we discuss shearing nondissipative expansionfree perfect
fluid with constant scalar curvature. For perfect fluid $\eta=0$
and hence the metric reduces to Eq.(\ref{44}). Using values of $A$
and $B$ from Eq.(\ref{44}) in the field equations
Eqs.(\ref{16})-(\ref{19}) along with Eq.(\ref{24}), it follows
that
\begin{eqnarray}\nonumber
-\frac{6\tau^2}{F_c^4C^6}+\frac{C^4\dot{C}''}{\dot{C}}+C^3(2C''
+\frac{8\dot{C}'C'}{\dot{C}})+10C'^2C^2-\frac{1}{2}\frac{f(R_c)}{F_c}
&=&\frac{8\pi\rho}{F_c},\\\nonumber
\frac{D_C\sigma}{3}+\frac{\sigma}{C}&=&0,\\\nonumber
\frac{\tau^2}{F_c^4\dot{C}C^5}(\frac{6\dot{C}}{C}-\frac{\dot{\tau}}{\tau})+
C^4(\frac{\dot{C}''}{
C}+\frac{4C''}{C}+6\frac{\dot{C}'C'}{\dot{C}C}+\frac{10C'^2}{C^2})+\frac{1}{2}
\frac{f(R_c)}{F_c} &=&\frac{8\pi p_r}{F_c},\\\nonumber
\frac{-\tau^2}{F_c^4\dot{C}C^5}(\frac{3\dot{C}}{C}-\frac{\dot{\tau}}{\tau})-
C^4(\frac{\dot{C}'C'}{\dot{C}
C}-\frac{3C'^2}{C^2}-\frac{C''}{C})+\frac{1}{C^2}+\frac{1}{2}\frac{f(R_c)}{F_c}
&=&\frac{8\pi p_{\perp}}{F_c}.
\end{eqnarray}
In this case, the Bianchi identities are reduced to the same
expression as in GR, i.e.,
\begin{eqnarray}\label{53}
&&\dot{\rho}+2(p_{\perp}-p_r)\frac{\dot{C}}{C}=0,\\\label{56}
&&p_r'+(\rho+p_r)\frac{\dot{C}'}{\dot{C}}+2(\rho+2p_r-p_{\perp})\frac{C'}{C}=0.
\end{eqnarray}
We would like to mention here that for isotropic fluid, i.e.,
$p_r=p_{\perp}$, Eq.(\ref{53}) implies that energy density $\rho$
depends only on $r$.

\subsection{Shearing Expansionfree Dust}

In this case, we have $p_r=0=p_{\perp}$ and hence Eq.(\ref{56})
becomes
\begin{equation}\label{54}
\frac{\dot{C}'}{\dot{C}}+2\frac{C'}{C}=0
\end{equation}
whose integration gives
\begin{equation}\label{55}
\dot{C}=\frac{g_2(t)}{C^2},\quad C'=\frac{g_3(r)}{C^2},
\end{equation}
where $g_2(t)$ and $g_3(r)$ are arbitrary functions of $t$ and $r$
respectively. Further, integrating Eq.(\ref{55}) with respect to the
corresponding arguments of the arbitrary functions, we get
\begin{equation}\label{58}
\frac{C^3}{3}=\int{g_2(t)}dt+\int{g_3(r)}dr
\end{equation}
which can be written as
\begin{equation}\label{59}
C^3=\psi(t)+\chi(r),
\end{equation}
where
\begin{equation}\label{60}
\psi(t)=3\int{g_2(t)}dt, \quad\chi(r)=3\int{g_3(r)}dr,
\end{equation}
In view of Eq.(\ref{55}) and taking $\tau(t)=g_2(t)$, the metric in
Eq.(\ref{44}) is reduced to
\begin{equation}\label{61}
ds^2=F_c^4dt^2-\frac{1}{C^4}dr^2-C^2(t,r)(d\theta^{2}+\sin^2\theta
d\phi^{2})
\end{equation}
while Eq.(\ref{22}) becomes
\begin{equation}\label{62}
U=\frac{\dot{C}}{F_c^2}.
\end{equation}
When we use the standard GR limit, i.e., $F_c=1$, Eqs.(\ref{61}) and
(\ref{62}) are reduced to the corresponding GR results.

\section{Summary}

This paper investigates the gravitational collapse of a
spherically symmetric star in $f(R)$ theory of gravity. The star
is made up of viscous anisotropic fluid distribution which is
dissipating energy in the form of heat flow, null radiations and
shearing viscosity. The objective of this work is to explore the
effects of DE which is generated by modifying EH action. We have
discussed the consequences in view of existing GR results.

It is concluded that the contribution of DE terms decreases the mass
of the collapsing fluid with the passage of time and hence prevents
the fluid to collapse. In dynamical equations, such curvature terms
appear oftentimes to affect the passive gravitational mass and rate
of collapse.

We have found some solutions by assuming that fluid has no
expansion, i.e., $\Theta=0$. Herrera \emph{et al.} \cite{H2} has
made a comprehensive discussion on the physical meaning of
expansionfree evolution of the fluid. According to them such an
assumption causes the formation of a vacuum cavity inside the
collapsing fluid. Using this condition, we have obtained general
metric for dissipative fluid which is further reduced to
nondissipative case. For nondissipative solution, a locally
anisotropic perfect fluid solution is obtained with the assumption
of constant scalar curvature. This is further reduced to the dust
case.

It is noted here that implication of hydrostatic equilibrium limit
yields the same expression as in GR \cite{H2}, i.e.,
\begin{equation}\setcounter{equation}{1}
D_Cp_r+2\frac{(p_r-p_\perp)}{C}=-\frac{(\rho+p_r)}{C(C-2M)}(M+4\pi
p_rC^3).
\end{equation}
The reason is that $f(R)$ term does not affect contracted Bianchi
identities which are used to obtain the above expression along
with the use of Eq.(\ref{53}) giving time independent $\rho$ in an
isotropic pressure case. This change in energy density can be
interpreted as the rate of work being done by the force of locally
anisotropic pressure. It would be worthwhile to investigate these
issues using other symmetries in $f(R)$ theory for the complete
understanding of gravitational collapse.

\vspace{0.5cm}

{\bf Acknowledgment}

\vspace{0.25cm}

We would like to thank the Higher Education Commission, Islamabad,
Pakistan for its financial support through the {\it Indigenous Ph.D.
5000 Fellowship Program Batch-III}.

\end{document}